\input{amssymb.sty}
\documentclass[aps,pre,twocolumn,amsmath,amssymb]{revtex4}
\usepackage{graphicx}
\usepackage{dcolumn}
\usepackage{bm}
\usepackage{epsfig}

\begin{document}
\title{Vectorial Loading of Processive Motor Proteins: Implementing a Landscape Picture}

\author{Young C.\ Kim}
\author{Michael E.\ Fisher}
\affiliation{Institute for Physical Science and Technology, University of Maryland, College Park, MD 20742, USA}

\date{\today}

\begin{abstract}

Individual processive molecular motors, of which conventional kinesin is the most studied quantitatively, move along polar molecular tracks and, by exerting a force ${\bm F} = (F_x,F_y,F_z)$ on a tether, drag cellular cargoes, {\em in vivo}, or spherical beads, {\em in vitro}, taking up to hundreds of nanometer-scale steps. From observations of velocities and the dispersion of displacements with time, under measured forces and controlled fuel supply (typically ATP), one may hope to obtain insight into the molecular motions undergone in the individual steps. In the simplest situation, the load force ${\bm F}$ may be regarded as a scalar resisting force, $F_x < 0$, acting parallel to the track: however, experiments, originally by Gittes {\em et al.} (1996), have imposed perpendicular (or vertical) loads, $F_z > 0$, while more recently Block and coworkers (2002, 2003) and Carter and Cross (2005) have studied {\em assisting} (or reverse) loads, $F_x > 0$, and also sideways (or transverse) loads $F_y \neq 0$.

We extend previous mechanochemical kinetic models by explicitly implementing a free-energy landscape picture in order to allow for the full vectorial nature of the force ${\bm F}$ transmitted by the tether. The load-dependence of the various forward and reverse transition rates is embodied in load distribution vectors, ${\bm \theta}_j^+$ and ${\bm \theta}_j^-$, which relate to {\em substeps} of the motor, and in next order, in compliance matrices ${\bm \eta}_j^+$ and ${\bm \eta}_j^-$. The approach is applied specifically to discuss the experiments of Howard and coworkers (1996) in which the buckling of partially clamped microtubules was measured under the action of bound kinesin molecules which induced determined perpendicular loads. But in the normal single-bead assay it also proves imperative to allow for $F_z > 0$: the appropriate analysis for kinesin, suggesting that the motor ``crouches" on binding ATP prior to stepping, is sketched. It yields an expression for the velocity, $V(F_x,F_z;\mbox{[ATP]})$, needed to address the buckling experiments.
\end{abstract}

\maketitle

\section{Introduction}
\label{sec1}

\begin{figure}[ht]
\vspace{-0.7in}
\centerline{\psfig{figure=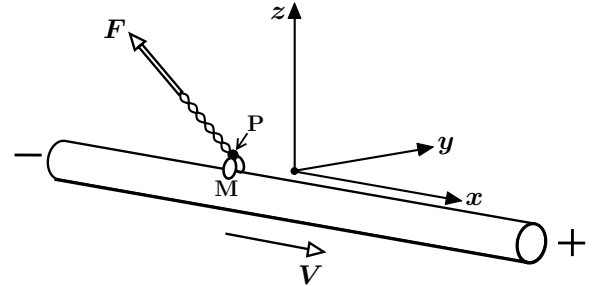,width=3.5in,angle=0}}
\vspace{-1.8in}
\caption{Schematic representation of a motor protein, M, bound to a molecular track (depicted as a microtubule with $+$ and $-$ ends). The point of attachment of the tether to the motor body is marked P. The force ${\bm F} = (F_x,F_y,F_z)$ denotes the tension transmitted by the tether. The cartesian coordinate system has its origin at the point of attachment of a stationary motor at a binding site further along the track. The sense of the mean velocity, $V$, under low loads is indicated by an arrow parallel to the track. \label{fig1}}
\end{figure}

A processive motor protein \cite{how} is an individual protein molecule that in an appropriate aqueous solution binds to a linear periodic directed molecular track and, when fueled via diffusion by suitable molecules (specifically ATP in cases of most interest), takes tens to hundreds of discrete steps along the track before dissociating. Such a motor steps overwhelmingly in a single characteristic direction, which we will identify as parallel to the positive $x$-axis: see Figure \ref{fig1}. 
Thus conventional kinesin walks towards the plus or fast-growing end of a microtubule while myosin V moves towards the plus or `barbed' end of an actin filament \cite{how}. In stepping along its track a motor exerts a tension ${\bm F}=(F_x,F_y,F_z)$ on a molecular tether the other end of which is, {\em in vivo}, bound to some cellular organelle or vesicle while for {\em in vitro} experiments it is firmly attached to a silica or plastic bead that may be controlled with the aid of an optical trap \cite{how}. For most purposes one may regard the molecular track as fixed in space: relatively rigid microtubules are typically clamped to a microscope slide while the more flexible actin filaments can be stretched between two further beads that are attached to the filament ends and held in place by individual optical traps \cite{how}.

In the conceptually most straightforward experiments \cite{how,koj:mut,vis:sch} a motor (M in Fig.\ \ref{fig1}) advances along the track in a positive direction in a solution of fixed fuel concentration, say [ATP], under a {\em resistive} load the $x$-component of which, $F_x < 0$, is observed or controlled. Measurements of the mean velocity and the dispersion \cite{vis:sch,sch:vis}:
 \begin{equation}
  V \approx \langle x(t)\rangle/t, \hspace{0.1in} \mbox{and} \hspace{0.1in} D \approx [\langle x^2(t)\rangle - \langle x(t)\rangle^2]/t,  \label{eq.1}
 \end{equation}
as functions of $F_x$ and [ATP], where $x(t)$ represents the displacement of the motor along the track at time $t$, may then be analyzed \cite{fis:kol,fis:kol2,kol:fis} with the aim of extracting details of the mechanism by which the motor takes individual steps. In particular one would like to identify and quantify any {\em substeps} or intermediate motions.

To investigate further experimentally it is rather natural to {\em impose} an oppositely directed {\em reverse} or {\em assisting} load on the motor so that $F_x > 0$. Indeed, such experiments have been performed on kinesin, first by Coppin {\em et al.} \cite{cop:pie}, and, more recently, by Block and coworkers \cite{lan:asb,blo:asb}, and by Carter and Cross \cite{car:cro}. In a preliminary computation to gain theoretical insight regarding the effects of reverse loading, Fisher and Kolomeisky ({\bf FK}) \cite{fis:kol2} examined the fairly noisy data of Coppin {\em et al.} \cite{cop:pie} for the mean velocity of kinesin under assisting loads up to $F_x {\,=\,} +\, 6$ pN. Merely by continuing the formulae developed for negative, i.e. resistive values of $F_x$, analytically through $F_x {\,=\,} 0$ to positive values, {\bf FK} obtained an apparently reasonable description of the observed `acceleration ratios,' $V(F_x {\,=\,} 5\; \mbox{pN})/V(F_x {\,=\,} 0)$. These varied from about 3.0 to 1.4 as [ATP] increased from 5 $\mu$M to 1 mM. The more recent experiments, however, do not confirm these results \cite{lan:asb,blo:asb,car:cro}; on the contrary, for [ATP] $\gtrsim 50 \,\,\mu$M the data indicate {\em no} acceleration or even a {\em de}celeration, for $F_x$ up to $10\,$-15 pN, i.e., $V(F_x {\,>\,} 0) \lesssim V(F_x {\,=\,} 0)$.

\begin{figure}[ht]
\vspace{-0.78in}
\centerline{\psfig{figure=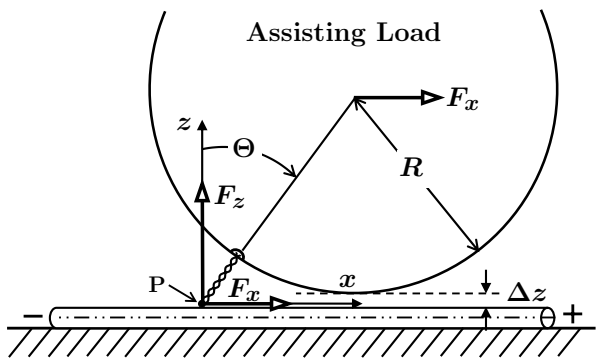,width=3.5in,angle=0}}
\vspace{-1.6in}
\caption{Depiction of the configuration of a bead, of radius $R$, and a tether, of length $l$, attached to the motor body at P when the motor (moving on a microtubule clamped to a substrate) is subjected to an {\em assisting}, $F_x > 0$, load. The offset $\Delta z$ denotes the root mean square thermal fluctuation of the bead as limited by collision with the track surface [12]. \label{fig2}}
\end{figure}

But, irrespective of the experimental observations, an examination of the geometry entailed in a standard single-bead assay with an assisting load, as illustrated in Figure \ref{fig2} \cite{svo:blo}, 
reveals that switching from a resistive to an assisting load should {\em not} be described merely by the change in sign of a scalar load force. Rather the true vectorial character of the force ${\bm F}$, acting at the {\em point of attachment}, P, of the tether to the body of the motor should be recognized. Even if sideways, $F_y \neq 0$, components of ${\bm F}$ may be neglected --- although Block and coworkers \cite{lan:asb,blo:asb} have imposed transverse loads --- a satisfactory description of the motor operation should seek to understand the mean velocity $V$, and likewise the dispersion $D$, as functions of $F_x$ {\em and} $F_z$.

Clearly, it would be advantageous experimentally to vary $F_z$ {\em independently} of $F_x$. With current optical trap technology this is not readily accomplished (since the traps are rather soft along the perpendicular or $z$-axis). To some degree, however, the issue can be addressed experimentally by employing beads of varying diameters. Thus, by reference to Figure \ref{fig2}, one sees that increasing the bead radius, $R$, increases $F_z$ relative to a controlled value of $F_x$ via
 \begin{equation}
  \frac{F_z}{F_x} = \cot\Theta = \frac{ R + \Delta z}{[(l - \Delta z)(2R + l + \Delta z)]^{1/2}} \simeq \sqrt{\frac{R}{2l}},  \label{eq.2}
 \end{equation}
\\in which $l$ is the length of the tether while $\Delta z^2$ represents the mean squared thermal fluctuation of the bead in the $z$ direction \cite{svo:blo}.

A decade ago, however, Howard stressed the vectorial nature of the load and the significance of independently observing the perpendicular or vertical component, $F_\perp \equiv F_z$, and measuring its effects on $V$. To this end Gittes, Meyh\"{o}fer, Baek and Howard ({\bf GMBH}) \cite{git:mey} devised an ingenious microtubule (MT) buckling experiment. 
\begin{figure}[hb]
\vspace{-0.78in}
\centerline{\psfig{figure=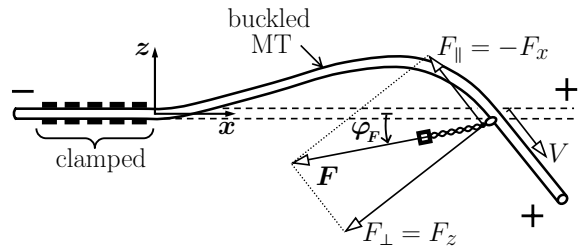,width=3.5in,angle=0}}
\vspace{-2.2in}
\caption{Arrangement of the microtubule (MT) buckling experiment [13] showing an originally straight MT (dashed lines) buckled by a kinesin motor that bound to the MT and proceeded to step towards the plus end at a velocity $V$ that depends on the varying force ${\bm F}$ transmitted by the tether. \label{fig3}}
\end{figure}
In their protocol the minus end of an MT is chemically attached to a glass substrate while the plus end of the MT can interact freely with a sparse field of kinesin motors tethered randomly to the same substrate. Then, as the free end of the MT diffusively wanders across the surface, a single kinesin molecule will occasionally encounter the MT and bind to it. In the presence of ATP it then proceeds to step along the MT towards the plus end: see Fig.\ \ref{fig3}. As the motor moves, it will eventually start to buckle the MT. By recording the buckling process and analyzing the shapes of the MT, the two components of the force, ${\bm F} = (F_x,F_z)$, in the plane of the substrate can be determined. Likewise the varying speed, $V(t)$, along the MT can be measured.  The experiment is not easy and the resulting data are quite noisy: nevertheless, {\bf GMBH} felt able to conclude that the vertical component, $F_z$, resulted in an {\em increase} in the overall mean speed.

In this article we extend the previous quantitative mechanochemical kinetic approach \cite{fis:kol,fis:kol2,kol:fis} in order to explicitly account for the vectorial character of the load force in analyzing the stochastic motions of motor proteins. The resulting formalism has been applied to the recent experiments of Block and coworkers on kinesin \cite{fis:kim,kim:fis}. Contrary to their model description, which implies sideways (or transverse) motions of the motor while stepping, we find no cause to invoke displacements of the point of attachment outside the $(x,z)$ plane. However, our analysis indicates that a kinesin motor `crouches' on binding ATP, that is, the point of attachment of the tether moves {\em downwards} towards the microtubule track, apparently by about 0.5 to 0.8 nm before rapidly completing a unitary `swing' or diffusive step of close to 8.0 nm transferring the motor to the next binding site on the track \cite{car:cro,nis:mut}.

Using the corresponding fitted kinetic parameters, we have revisited the Gittes {\em et al.} microtubule buckling experiment \cite{git:mey}. As we explain below, our analysis avoids a simplifying assumption made in their discussion and seems qualitatively consistent with the buckling data; however, it indicates that $V(F_x,F_z;\mbox{[ATP]})$ generally {\em decreases} when $F_z$ increases (at fixed $F_x$ and [ATP]) rather than increases as {\bf GMBH} argued. Nevertheless, one may note that in the buckling experiments the motor moves progressively away from a curved and, hence, stressed region of the microtubule. This raises the interesting question as to the degree to which the {\em curvature} of a microtubule protofilament might affect the motility of kinesin.

\section{Intermediate States and Substeps}
\label{sec2}

The modeling of motor protein motility may be carried out at different levels \cite{how,lei:hus,jul:adj,kol:wid,bus:kel}. For concreteness and relative simplicity, we will focus on motors powered by the hydrolysis of ATP and moving like kinesin via steps of fixed size. Then there are specific binding sites located at positions $x = ld$ $(l=0,\pm 1, \pm 2,\cdots)$ along the periodic track where $d$ is the step distance corresponding to the track periodicity. For a microtubule one has $d \simeq 8.2$ nm, representing the size of a tubulin dimer. In the absence of molecular fuel, $\mbox{[ATP]} = 0$, the motor will be bound at a specific site $l$ in a nucleotide-free state to be labeled $0_l$. We will overlook the retention of ADP in the weakly bound head of kinesin \cite{how,sch:cla} and neglect the rates of spontaneous dissociation from the track: but see \cite{fis:kol2,kol:fis2}. 

When, accepting the evidence for ``tight coupling'' \cite{how}, the motor translocates from site $l$ to the adjacent site $l+1$, by processing a single fuel molecule, it undergoes, in the simplest linear reaction sequence considered here, a series of $N$ intermediate mechanochemical transitions from states $j=0_l$ to $1_l$ to $\cdots$ to $(N-1)_l$ to $N_l\equiv 0_{l+1}$. When a motor is powered by the hydrolysis of ATP, the biochemical evidence indicates that $N=4$ distinct enzymatic states should be recognized \cite{how}; however, the degree to which these are linked to significantly different {\em mechanical} states is a significant object of investigation. Clearly, the simplest nontrivial model requires $N=2$ states: the transition from state $0_l$ to $1_l$ then corresponds to the binding of ATP which is followed by the hydrolysis process that is completed, with loss of ADP and P$_{\mbox{\scriptsize i}}$, when the motor moves on to state $2_l\equiv 0_{l+1}$. Again for simplicity, we overlook the coupling between two distinct heads of a motor as entailed in the so-called hand-over-hand motion now well established for kinesin: see, e.g. \cite{sch:cla,yil:tom,klu:hoe:gil,kol:phi}. Nevertheless, this can be accommodated formally simply by doubling the nominal step size $d$ and allowing for twice as many intermediate states to describe periodic double-steps. (See, e.g., the analysis used in \cite{kol:fis} for myosin V which, indeed, exhibits steps of fluctuating size.)

In the simplest, traditional chemical kinetic pictures \cite{fis:kol,lei:hus} one introduces transition rates, forwards and backwards, between states along a reaction pathway: thus we will associate rates $u_j$ and $w_j$ with the transitions from state $j_l$ forward to state $(j$$\,+\,$$1)_l$ and backward to state $(j$$\,-\,$$1)_l$, respectively. Note that owing to the periodicity of the stepping process, the rates $u_j$ and $w_j$ are independent of $l$.

However, to describe the action of motor proteins under variable loads it is essential to describe the dependence of the various rates on the loads imposed, or, complementarily, on the stresses induced: and how the influence of the load is distributed over the different mechanochemcial transitions should be a prime question \cite{fis:kol}. Since purely chemical steps are typically very fast on the scale of the overall mechanical movements, a rather basic theoretical picture \cite{jul:adj} postulates a distinct free energy surface, depending on molecular coordinates, for each separate mechanically fluctuating biochemical state. Motions occur by diffusion through the multidimensional configurational space while the probability of a formally instantaneous ``chemical jump'' from one surface to another varies with the configuration. One may argue, however \cite{kol:wid}, that for practical purposes this scheme may be reduced to an effective {\em mechanochemical description} in which points along the traditional chemical ``reaction coordinate'' are, for a motor protein, simply identified with specific positions of the motor along the linear track.

By this route, as previously adopted \cite{fis:kol2,kol:fis}, one may, indeed, hope to identify {\em substeps}, $d_0$, $d_1$, $\cdots$, in which the motor moves from a binding site, say, at $x_0$ to the next site $x_0 + d$ via intermediate locations $x_1=x_0 + d_0$, $x_2 = x_0 + d_0 + d_1$, $\cdots$, with $\sum_{j=0}^{N-1}d_j = d$. Likewise, within the chemical picture, one may identify successive, unstable ``transition states,'' say $j_l^+ \equiv j_{l+1}^-$, along the reaction coordinate each lying {\em between} the locally stable intermediate states $j_l$ and $j_{l+1}$. In mapping the reaction coordinate onto the track, one may then decompose a substep $d_j$ from $x_j$ to $x_{j+1}$ into the sum $(d_j^+ + d_{j+1}^-)$ thereby locating the transition state at $x_j^+ = x_j+d_j^+ = x_{j+1}-d_{j+1}^- \equiv x_{j+1}^-$: see Fig.\ \ref{fig1} of \cite{fis:kol2} for depictions of such mappings with $N=2$ and $N=4$, as fitted to data for kinesin under resisting $(F_x < 0)$ loads \cite{vis:sch,sch:vis}.

It is rather clear that in such a treatment, in which the reaction pathway is assumed to be parallel to the $x$-axis, only the longitudinal component, $F_x$, of the load will be coupled to displacements of the motor and so affect the rate constants. However, the initial substep predicted by this approach $(d_0 = 1.8\,$-2.1 nm$)$ appears to be inconsistent with high-resolution spatio-temporal observations of individual and averaged steps \cite{fis:kim,kim:fis,nis:mut,car:cro}. Furthermore, this approach has failed to account satisfactorily for velocity measurements of kinesin under reverse loading \cite{lan:asb,blo:asb,car:cro}. Accordingly, apart from general theoretical considerations, it seems important to extend the previous discussions.

\begin{figure}[ht]
\vspace{-0.78in}
\centerline{\psfig{figure=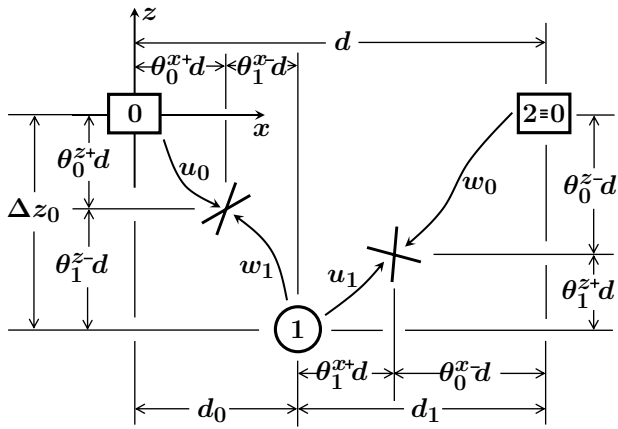,width=3.5in,angle=0}}
\vspace{-1.6in}
\caption{Schematic depiction within an $(N{\,=\,}2)$-state model of the motion in the $(x,z)$ plane of the point of attachment, $\mbox{P}{\,\equiv\,} {\bf r} {\,=\,} (x,y,z)$, of the tether to the motor body. Two bound nucleotide-free states, labeled $0$ and $0{\,\equiv\,} 2$, are shown together with a single $(N{\,-\,}1{\,=\,}1)$ intermediate mechanochemical state, $1$, and the two $(N{\,=\,}2)$ corresponding transition states ($0^+{\,\equiv\,}1^-$ and $1^+{\,\equiv\,}2^-$). See further discussion in the text. \label{fig4}}
\end{figure}

Evidently, a significant next step in considering the load-dependence is to recognize the vectorial character of the imposed force by allowing states along the reaction pathway to be mapped onto movements of the motor within the full three-dimensional space of the track, motor, tether, and cargo: see Fig.\ \ref{fig1}. More concretely we may hope to follow the course of the point of attachment P $\Rightarrow {\bm r}(t) = [x(t),y(t),z(t)]$ of the tether to the motor as steps are taken: see Figs.\ \ref{fig1} and \ref{fig2}. As varying values, $z(t)$ and $y(t)$, arise one can view the motor as moving up or down or swinging from side to side when it walks along the track from a binding site $l$ to the next binding site $l+1$. The simplest ($N{\,=\,}2$)-state model embodying this concept is illustrated in Fig.\ \ref{fig4}. 
It has been supposed, first, that the motion of P may be regarded as confined to the $(x,z)$ plane, i.e., that sideways (or transverse) motions may be neglected. Of course this should be subject to experimental test: but for kinesin \cite{lan:asb,blo:asb,kim:fis} it proves quite adequate \cite{kim:fis}.

Then in Fig.\ \ref{fig4} the rectangular boxes labeled ``0'' and ``$2\equiv 0$'' represent bound, nucleotide-free states at successive sites $l$ and $l+1$, along the track. The circle labeled ``1'' denotes the ATP bound state; this state may also include subsequent ADP-P$_{\mbox{\scriptsize i}}$ and ADP biochemical states. This $N{\,=\,}2$ level model, however, cannot distinguish situations in which the hydrolysis, ADP, and P$_{\mbox{\scriptsize i}}$-release chemical reactions take place physically at (or close to) the mechanical state ``1'' or close to ``2''. (And recall also the remark above concerning hand-over-hand motion.)

The (unlabeled) crosses in Fig.\ \ref{fig4} represent the location of the two transition states, $0^+ \equiv 1^-$ and $1^+\equiv 2^-$. Also introduced in the figure are the dimensionless {\em load distribution factors} $\,\,\theta_{j}^{x+}$, $\theta_{j}^{x-}$, $\theta_{j}^{z+}$ and $\theta_{j}^{z-}$ (following \cite{fis:kol,fis:kol2}). These turn out to be crucial fitting parameters in describing experimental data. Evidently they serve to specify the substeps via
 \begin{equation}
  d_j^\pm = \theta_{j}^{x\pm}d \hspace{0.2in} \mbox{and} \hspace{0.2in} d_j = (\theta_{j}^{x+} + \theta_{j+1}^{x-})d, \label{eq.3}
 \end{equation}
while vertical (or perpendicular) displacements of the point of attachment are specified by
 \begin{equation}
  \Delta z_j^\pm = \theta_{j}^{z\pm}d \hspace{0.2in} \mbox{and} \hspace{0.2in} \Delta z_j = (\theta_{j}^{z+} + \theta_{j+1}^{z-})d,  \label{eq.4}
 \end{equation}
with, of course, similar expression for $\Delta y^\pm$ in terms of $\theta_{j}^{y\pm}$. Since the stepping is periodic one must finally have
 \begin{equation}
  \sum_{j=0}^{N-1} (\theta_{j}^{y+} + \theta_{j}^{y-}) = \sum_{j=0}^{N-1} (\theta_{j}^{z+} + \theta_{j}^{z-}) = 0. \label{eq.5}
 \end{equation}

\section{Load-dependence of Transition Rates}
\label{sec3}

Now, under any external load, ${\bm F}$, the rate constants, $u_j$ and $w_j$, must change. But how? To answer, let us accpet a set of ``free-energy landscapes'', for distinct biochemical states each depending on some set of molecular configurations \cite{jul:adj}. Then one can contemplate a ``reduced landscape'' in which the set of distinct landscapes is, in effect, combined by identifying a traditional reaction coordinate, while at the same time retaining a particular {\em mechanical coordinate}, like the displacement, $x$, of the motor along its track \cite{bus:kel}. (One may, incidently, note a critique \cite{fis} of this general approach.) It is natural in the present case, as we have demonstrated, to extend the retained mechanical coordinate to be the vector ${\bm r}$ specifying, in real space, the point of attachment, P, of the tether to the motor. Then, if one supposes the reaction pathway can, at least on average, be mapped on to the motion of P, one may dispense with any explicit consideration of the reaction coordinate. Thus we are led to postulate a more-or-less well defined, presumably smooth and differentiable free-energy function, say $\Phi({\bm r})$, which in the absence of a load (i.e., ${\bm F} = 0$) respects the periodicity of the track so that
 \begin{equation}
  \Phi ({\bm r}) = \Phi ({\bm r} + d\hat{\bm x}),  \label{eq.6}
 \end{equation}
where $\hat{\bm x}$ is a unit vector parallel to the $x$ axis. (See also \cite{how} Chaps 15, 16, etc., and Sachs and Lecar \cite{sac:lec}.)

The various mechanochemical states $0_l$, $1_l$, $\cdots$, $j_l$, $\cdots$ are then to be identified with corresponding valleys or potential wells, i.e., minima in $\Phi ({\bm r})$, located, say at ${\bm r}_j + ld\hat{\bm x}$ for $j=0,1,\cdots$ (mod $N$) and $l=0,\pm 1,\pm 2,\cdots$. Following the traditional chemical picture, successive valleys along the reaction path will be linked via free energy barriers corresponding to the respective transition states. These, in turn, will be represented, by cols (or passes or saddles) in $\Phi ({\bm r})$ at points ${\bm r}_j^+ \equiv {\bm r}_{j+1}^-$.

When a force ${\bm F}$ is exerted on the motor's tether the (now reduced) free energy landscape must be deformed. It is then most reasonable (but surely not {\em fully} `realistic') to suppose that
 \begin{equation}
  \Phi ({\bm r};{\bm F}) = \Phi({\bm r}) - {\bm F}{\bm\cdot\,}{\bm r}.   \label{eq.7}
 \end{equation}

Now, once again in accord with traditional chemical reaction rate theories (see, e.g.\ \cite{how}), the rate $u_j$ of the forward reaction from a state $j$ to $j+1$ will be dominated by the Boltzmann factor for overcoming the corresponding barrier; and the same goes for the backwards reaction to $j-1$. Thus, if $\Phi_j({\bm F})$ is the free energy level at the bottom of the $j$th valley while $\Phi_j^+({\bm F})$ and $\Phi_j^-({\bm F})$ are the height of the associated forward and rearward col, respectively, we may suppose
 \begin{eqnarray}
 u_j({\bm F}) & \propto & e^{-[\Phi_j^+({\bm F})-\Phi_j({\bm F})]/k_{\mbox{\tiny B}}T}, \nonumber \\
 w_j({\bm F}) & \propto & e^{-[\Phi_j^-({\bm F})-\Phi_j({\bm F})]/k_{\mbox{\tiny B}}T}. \label{eq.8}
 \end{eqnarray}

To proceed so as to obtain the force-dependence of the rates up to quadratic order in ${\bm F}$ on the basis of this landscape picture, consider first the forward rate constant $u_j$. In the absence of the load the corresponding valley is located at ${\bm r}_j$ while the appropriate col (describing the transition state $j^+$) is at ${\bm r}_j^+ \equiv {\bm r}_j + {\bm \theta}_j^+d$ where we have introduced the {\em forward load distribution vector} $\;{\bm \theta}_j^+ = (\theta_{j}^{x+},\theta_{j}^{y+},\theta_{j}^{z+})$. Expansion of the free energy in the valley then yields
 \begin{equation}
  \Phi({\bm r}) = \Phi_{j} + \mbox{$\frac{1}{2}$}({\bm r}-{\bm r}_{j}){\bm\cdot}\mbox{\bf A}_{j}{\bm\cdot}({\bm r}-{\bm r}_{j}) + O(|{\bm r}-{\bm r}_j|^{3}),  \label{eq.9}
 \end{equation}
where $\Phi_j \equiv \Phi({\bm r}_j)$ while {\bf A}$_j$ is a positive definite symmetric $3$$\,\times\,$$3$ matrix with elements $A_j^{\lambda\mu} \equiv (\partial^2\Phi/\partial\lambda\partial\mu)$ evaluated at ${\bm r}_j$, where $\lambda,\mu=x,y,z$. On the other hand, one may expand the free energy around the col in the form
 \begin{equation}
  \Phi({\bm r}) = \Phi_j^+ + \mbox{$\frac{1}{2}$} ({\bm r}-{\bm r}_j^+){\bm\cdot}\mbox{\bf A}_j^+ {\bm\cdot} ({\bm r}-{\bm r}_j^+) + \cdots,  \label{eq.10}
 \end{equation}
where $\Phi_j^+ \equiv \Phi({\bm r}_j^+)$ and {\bf A}$_j^+$ is again a $3$$\,\times\,$$3$ matrix with elements $(\partial^2\Phi/\partial\lambda\partial\mu)$ but now evaluated at ${\bm r}_j^+$. Note that by the character of a col or saddle point, the matrix {\bf A}$_j^+$, which is identical to {\bf A}$_{j+1}^-$, must be indefinite with at least one negative eigenvalue. The corresponding eigenvector identifies the optimal direction of the reaction, projected into ${\bm r}$-space as the system enters and leaves the transition state. It will be appropriate for us to assume that the remaining eigenvalues are positive.

The rate constant $u_j^0$ for ${\bm F}=0$ is then proportional to $\;\exp(-\Delta\Phi_j^{+0}/k_{\mbox{\scriptsize B}}T)$ where $\Delta\Phi_j^{+0}=\Phi({\bm r}_j^+)-\Phi({\bm r}_j)$ is the barrier height. Under a vectorial load ${\bm F}$, however, the positions of both valley and col shift, the changes being proportional to ${\bm F}$ in leading order. The new positions may be found by solving the equations $\nabla\Phi({\bm r};{\bm F})=0$ using (\ref{eq.9}), (\ref{eq.10}) and (\ref{eq.7}). Thus one finds the free energy difference between the shifted valley and col to be
 \begin{eqnarray}
  \Delta\Phi_j^+({\bm F}) & = & \Phi_j^+({\bm F}) - \Phi_j({\bm F}), \nonumber \\
  & = & \Delta\Phi_j^{+0} - {\bm F}{\bm\cdot} ({\bm r}_j^+ - {\bm r}_j) - \mbox{$\frac{1}{2}$}{\bm F}{\bm\cdot} {\bm \eta}_j^+ {\bm\cdot} {\bm F} + O(F^3), \nonumber \\  \label{eq.11}
 \end{eqnarray}
in which appears the matrix
 \begin{equation}
  {\bm \eta}_j^+ = (\mbox{\bf A}_j^+)^{-1} - (\mbox{\bf A}_j)^{-1},  \label{eq.12}
 \end{equation}
where by our construction, the inverse matrices of {\bf A}$_j$ and {\bf A}$_j^+$ are well defined. Note, again, that ${\bm r}_j^+ - {\bm r}_j = {\bm \theta}_j^+ d$.

By the same arguments, the free energy barrier for the reverse rate $w_j$ is determined by the transition state $j^-$ located at ${\bm r}_j^- = {\bm r}_j - {\bm \theta}_j^- d$ and is given by
 \begin{eqnarray}
  \Delta\Phi_j^- ({\bm F}) & = & \Phi_j^- ({\bm F}) - \Phi_j ({\bm F}), \nonumber \\
  & = & \Delta\Phi_j^{-0} + {\bm F}{\bm\cdot} {\bm \theta}_j^- d + \mbox{$\frac{1}{2}$} {\bm F}{\bm\cdot} {\bm \eta}_j^- {\bm\cdot} {\bm F} + O(F^3), \nonumber \\ \label{eq.13}
 \end{eqnarray}
where ${\bm \theta}_j^- = (\theta_{j}^{x-},\theta_{j}^{y-},\theta_{j}^{z-})$ is the reverse load distribution vector while
 \begin{equation}
  {\bm \eta}_j^- = (\mbox{\bf A}_j^-)^{-1} - (\mbox{\bf A}_j)^{-1},  \label{eq.14} 
 \end{equation}
in which $\mbox{\bf A}_j^- \equiv \mbox{\bf A}_{j-1}^+$ is the matrix of the second derivatives of $\Phi$ evaluated at ${\bm r}_j^- \equiv {\bm r}_{j-1}^+$.

Finally, we may express the load-dependence of the rate constants as
 \begin{eqnarray}
  u_j({\bm F}) & = & u_j^0 \exp\{+[{\bm \theta}_j^+ {\bm\cdot} {\bm F}d + \mbox{$\frac{1}{2}$}{\bm F} {\bm\cdot} {\bm \eta}_j^+ {\bm\cdot} {\bm F} + O(F^3)]/k_{\mbox{\scriptsize B}}T\}, \nonumber \\ \label{eq.15} \\
  w_j({\bm F}) & = & w_j^0 \exp\{-[{\bm \theta}_j^- {\bm\cdot} {\bm F}d + \mbox{$\frac{1}{2}$}{\bm F} {\bm\cdot} {\bm \eta}_j^- {\bm\cdot} {\bm F} + O(F^3)]/k_{\mbox{\scriptsize B}}T\}.  \nonumber \\ \label{eq.16}
 \end{eqnarray}
Evidently, the load distribution vectors ${\bm \theta}_j^+$ and ${\bm \theta}_j^-$ simply generalize the previous load distribution scalars \cite{fis:kol,fis:kol2,kol:fis} and serve to locate the intermediate mechanochemical states and determine the vectorial character of the substeps linking them. By the underlying periodicity they must satisfy
 \begin{equation}
  \sum_{j=0}^{N-1} ({\bm \theta}_j^+ + {\bm \theta}_j^- ) = \hat{\bm x}, \label{eq.17}
 \end{equation}
which simply summarizes (\ref{eq.3}) and (\ref{eq.5}).

Similarly, the matrices ${\bm \eta}_j^+$ and ${\bm \eta}_j^-$ serve to measure the relative compliances of the various transition states. We may note that, by (\ref{eq.12}) and (\ref{eq.14}), the consecutive differences
 \begin{equation}
  {\bm \eta}_j^+ - {\bm \eta}_{j+1}^- = (\mbox{\bf A}_{j+1})^{-1} - (\mbox{\bf A}_j)^{-1},  \label{eq.18}
 \end{equation}
are independent of the properties of the transition states; then, via periodicity, one concludes that
 \begin{equation}
  \sum_{j=0}^{N-1} ({\bm \eta}_j^+ - {\bm \eta}_j^- ) = 0.  \label{eq.19}
 \end{equation}

In summary, one might be tempted to regard the load-dependence expressions (\ref{eq.15}) and (\ref{eq.16}) merely as phenomenological expansions in powers of ${\bm F}d/k_{\mbox{\scriptsize B}}T$; however, the crucial feature lies in the physical interpretation of the load distribution vectors ${\bm \theta}_j^\pm$ and the compliance matrices ${\bm \eta}_j^\pm$ which, in turn, yields the constraints embodied in (\ref{eq.17}) and (\ref{eq.19}).

\section{Motility of Kinesin under Vectorial Loading}
\label{sec4}

Our primary aim now, as an application of the foregoing analysis, is to reconsider the {\bf GMBH} buckling experiment \cite{git:mey}. As explained in the Introduction with the aid of Fig.\ \ref{fig3}, the crucial feature of that study was the direct determination of the perpendicular force $F_\perp \equiv F_z$ and, independently, the longitudinal or parallel component, given by $F_\parallel \equiv - F_x$ (since the load is always resistive from the perspective of the track). We will suppose, in the absence of contrary evidence, that the microtubule (MT) does not twist and that the kinesin, when it attaches to the MT, binds on to the nearest protofilament which, again we suppose for simplicity, does not spiral around the MT \cite{how}. Then any externally generated transverse force components may be neglected: i.e., $F_y = 0$.

In order to analyze the buckling data theoretically following {\bf GMBH}, one needs an explicit expression for $V(F_x,F_z;\mbox{[ATP]})$, the velocity of the motor along the track as a function of $F_x$, $F_z$ at fixed [ATP]. In default of a better description, {\bf GMBH} postulated a simple linear dependence of $V$ on $F_x$ and $F_z$. However, we may hope to do better by using the recent data of Block and coworkers \cite{lan:asb,blo:asb} ({\bf BASL}) who imposed assisting and resisting (as well as sideways) loads. The {\bf BASL} experiments studied squid kinesin whereas {\bf GMBH} employed kinesin from bovine brains: undoubtedly this and other disparate experimental details should play some role. Nevertheless, experience suggests that the quantitative effects should not be great.

\begin{figure}[ht]
\vspace{-0.9in}
\centerline{\psfig{figure=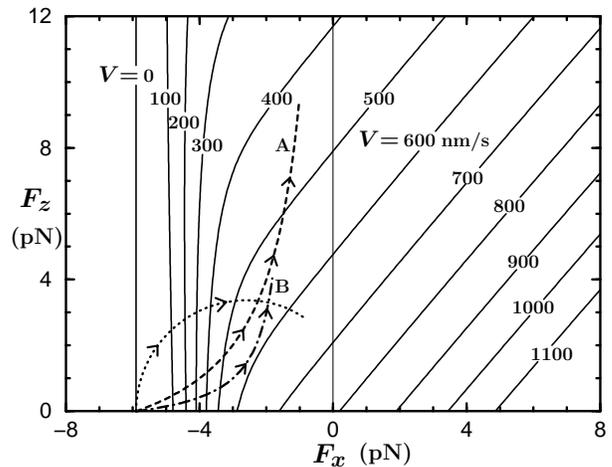,width=3.5in,angle=0}}
\vspace{-1.2in}
\caption{Contours of the velocity, $V(F_x,F_z;\mbox{[ATP]})$, of kinesin at saturating ATP in the $(F_x,F_z)$ plane as derived from the data of Block {\em et al.} \cite{blo:asb,kim:fis,fis:kim2}. The dotted, dashed and dot-dashed loci, related to the microtubule buckling experiment of Howard and coworkers \cite{git:mey} (see Fig.\ \ref{fig3}), are discussed in the text. \label{fig5}}
\end{figure}

Accordingly, in Fig.\ \ref{fig5} 
we present two-dimensional velocity-contour plots of conventional kinesin in the $(F_x,F_z)$ plane at saturating ATP concentration by using the simple $(N$$\, = \,$$2)$-state kinetic model as fitted to the {\bf BASL} data. Details of our analysis will be discussed elsewhere \cite{kim:fis,fis:kim2}. For completeness, however, we explain briefly here how these contour plots have been derived.

In terms of the general $N{\,=\,}2$ expression \cite{fis:kol}
 \begin{equation}
  V = d(u_0 u_1 - w_0 w_1)/(u_0 + u_1 + w_0 + w_1),  \label{eq.20}
 \end{equation}
the dependence of the velocity as a function of [ATP] under (effectively) {\em zero load} may be accounted for, following \cite{fis:kol2}, by
 \begin{equation}
  u_0^0 = k_0^0 \mbox{[ATP]}, \hspace{0.1in} w_0^0 = k_0^\prime \mbox{[ATP]}/(1+\mbox{[ATP]}/c_0)^{1/2},  \label{eq.21}
 \end{equation}
in which the final, square-root factor (introduced to take account of an ATP regeneration system and a related trend in the stall force) plays only a small role. Then, with $d=8.2$ nm a good description of the kinesin data of {\bf BASL} is given by the rates
 \begin{eqnarray}
  k_0^0 & = & 1.35 \;\mu\mbox{M}^{-1}\mbox{s}^{-1}, \hspace{0.2in} w_1^0 = 5.0 \;\mbox{s}^{-1}, \hspace{0.2in} u_1^0 = 100 \;\mbox{s}^{-1}, \nonumber \\
  k_0^\prime & = & 2.04 \times 10^{-3} \;\mu\mbox{M}^{-1}\mbox{s}^{-1}, \hspace{0.2in} c_0 = 20 \;\mu\mbox{M},  \label{eq.22}
 \end{eqnarray}
which, apart for the larger value of $k_0^\prime$, are quite comparable to the original fits \cite{fis:kol2} based on the data of Visscher and coworkers \cite{vis:sch,sch:vis} (restricted to $F_x \leq 0$).

Now Block {\em et al.} \cite{blo:asb} measured the velocity $V$ only as a function of the parallel component $F_x$ (even though for both $F_x < 0$ and $F_x > 0$). However, in order to use (\ref{eq.20}) with the ${\bm F}$-dependence expressions (\ref{eq.15}) and (\ref{eq.16}), one also needs to know the perpendicular component $F_z$. The route to unraveling this puzzle depends, as indicated in the Introduction, on a consideration of the geometry of the bead-tether-motor-track configuration as illustrated in Fig.\ \ref{fig2}. However, further properties of the motor and the experimental set-up may need to be accounted for. Concretely, we desire a formula, say $F_z = {\mathcal F}_z(F_x)$, that correlates the imposed (but unmeasured) perpendicular component $F_z$ with the observed parallel component $F_x$ \cite{fis:kim}.

The most basic hypothesis --- termed Model 0 \cite{kim:fis,fis:kim2} --- is to suppose, following Fig.\ \ref{fig2}, that the tether angle $\Theta$ simply switches sign when $F_x$ passes through zero: this is expressed by
 \begin{equation}
  \mbox{Model 0:\hspace{0.4in}} F_z = F_x \cot\Theta(F_x) = c_\parallel |F_x|,  \label{eq.23}
 \end{equation}
where $c_\parallel = |\cot\Theta|$ follows from (\ref{eq.2}). Reasonable values for beads of diameter $2R = 0.50$ $\mu$m, a fluctuating scale $\Delta z = 5 $ nm \cite{svo:blo} and a tether length $l=60$ nm yield $\Theta \simeq 35^\circ$ and $c_\parallel \simeq 1.44$.

In practice, the thermal fluctuations already recognized by allowing for $\Delta z$ in Fig.\ \ref{fig2}, come into play strongly when $|F_x|\lesssim 0.5$ pN. Accordingly, a more realistic hypothesis (which is tested further in Sec.\ \ref{sec6} below) is embodied in the smoothed form
 \begin{equation}
 \mbox{Model I:\hspace{0.4in}} F_z = {\mathcal F}_z(F_x) = c_\parallel\sqrt{F_x^2 + F_0^2},  \label{eq.24}
 \end{equation}
in which the fluctuation amplitude $F_0 = 0.3$ pN proves appropriate in light of observed force fluctuations \cite{lan:asb,blo:asb}. Study of the {\bf BASL} data, however, reveals an unexpected and strong asymmetry in the relation $F_z = {\mathcal F}_z (F_x)$ for kinesin \cite{kim:fis,fis:kim2}. This can be represented effectively via an additive term as
 \begin{equation}
 \mbox{Model II:\hspace{0.in}} F_z = c_\parallel \left[\sqrt{F_x^2 + F_0^2} + \frac{F_1}{2}\left( 1+ \frac{F_x}{\sqrt{F_x^2 + F_0^2}}\right) \right]. \label{eq.25}
 \end{equation}
The amplitude $F_1$, which is found to be close to 2.0 pN, measures the asymmetry. Although the new term may appear as an artificial construct it can be interpreted rather directly in mechanical terms \cite{kim:fis,fis:kim2}. Note, incidentally, that the factor in large parentheses merely represents a smoothed Heaviside step function vanishing rapidly for $F_x < 0$.

On this basis, successful fits to the {\bf BASL} data [including measurements of the randomness $r(F_x;\mbox{[ATP]}) = D/dV$] are achieved with the load distribution vectors
 \begin{eqnarray}
  {\bm \theta}_0^+ & = & (0.120,\;0,\;-0.043), \hspace{0.1in} {\bm \theta}_1^+ = (0.032,\; 0,\; -0.026),  \nonumber \\
  {\bm \theta}_0^- & = & (0.950,\; 0,\; 0.100), \hspace{0.1in} {\bm \theta}_1^- = (-0.102,\; 0,\; -0.031). \nonumber \\ \label{eq.26}
 \end{eqnarray}
Of course, these satisfy (\ref{eq.17}) so that only six independent fitting parameters are entailed. The compliance matrices, ${\bm \eta}_0^+$ and ${\bm \eta}_0^-$, do not need to be invoked although one might reasonably presume that they are not totally negligible in reality.

Now we may note, using (\ref{eq.3}) and (\ref{eq.4}), that these load distribution vectors imply a very small forward step of only $d_0 \simeq 0.1$ to 0.2 nm (allowing for the fitting uncertainties) on binding ATP but a much larger {\em downwards} displacement, namely $\Delta z_0 = -0.5$ to $-0.8$ nm. In this sense, then, the motor appears to ``crouch'' before it completes the main forward step of magnitude $d_1\simeq 8.0$ to $8.1$ nm. It should also be remarked \cite{fis:kol2,kol:fis}, that the relatively large positively directed value of ${\bm \theta}_0^-$ means that it is the reverse rate $w_0$ that changes most under a resisting load and thereby leads to the motor stalling $(V=0)$.

With the parameters presented above, the contours plotted in Fig.\ \ref{fig5} follow from (\ref{eq.20}). We may remark that when the level of ATP is lowered the form of the contours changes little qualitatively but the scale is reduced in accord with (\ref{eq.20})-(\ref{eq.22}) which imply close-to Michaelis-Menton variation of $V(\mbox{[ATP]})$ at zero load. It should also be noted that the direct sampling of the $(F_x,F_z)$ plane by the Block and coworkers experiments \cite{vis:sch,sch:vis,lan:asb,blo:asb} is essentially confined to the locus specified by the $F_z = {\mathcal F}_z(F_x)$ relation and so, by Model I, is roughly given by $F_z = c_\parallel |F_x| \geq 0.3$ pN. In principle, the behavior further from this locus could be somewhat different than portrayed. In particular, the fact that the fitting was achieved without explicitly employing any compliance coefficients is responsible for the steep rise of the velocity contours in Fig.\ \ref{fig5} as the stall force, ${\bm F}_{\mbox{\scriptsize S}} \simeq - 5.9\, \hat{\bm x}$ pN, is approached. Thus, in the absence of the compliance matrices and further ${\bm F}$-dependent corrections, the stall condition $V=0$ combined with the periodicity constraint (\ref{eq.17}) for the load distribution vectors implies independence of $F_y$ and $F_z$. Some evidence on the likely magnitudes of the compliance effects is available from analysis of the transverse loading experiments of Block {\em et al.} \cite{blo:asb,fis:kim2}. Shifts in the contours, away from the ${\cal F}_z(F_x)$ locus, as large at 10 to $25 \%$ might be realized.

\section{Buckling of a Microtubule by a Motor}
\label{sec5}

\begin{figure}[ht]
\vspace{-0.2in}
\centerline{\psfig{figure=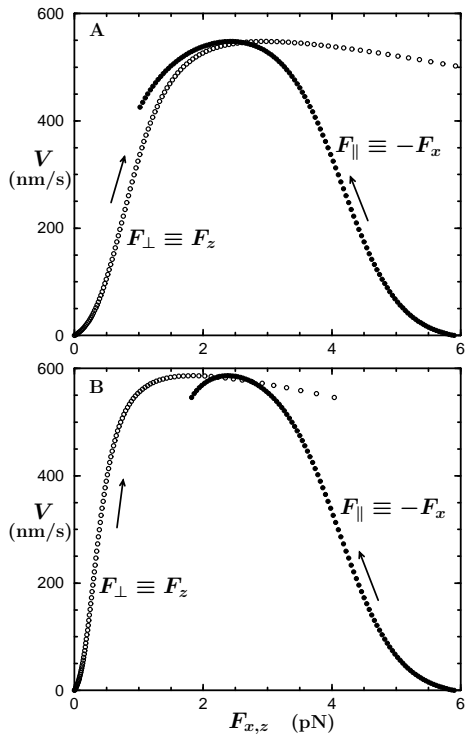,width=3.5in,angle=0}}
\vspace{-0.5in}
\caption{Velocity-force plots mimicking those observed in the microtubule buckling experiments of Howard and coworkers \cite{git:mey}. These plots are based on the putative force trajectories, A and B, respectively, shown in Fig.\ \ref{fig5}. The open circles are for the perpendicular force components, $F_\perp \equiv F_z$, the solid circles for the parallel components, $F_\parallel \equiv -F_x$. \label{fig6}}
\end{figure}

As explained above, in the experiments of {\bf GMBH} \cite{git:mey} the minus end of a microtubule was clamped to a flat glass substrate on which kinesin molecules were sparsely tethered. When an individual motor encountered and bound to the thermally flexing MT, it moved towards the plus end and soon buckled the MT: see Fig.\ \ref{fig3}. Then in time sequence (at 30 frames per second), the successive shapes of the MT were recorded. From the progress of the motor along the contour of the flexing MT the time varying velocity, $V(t)$, could be found by fitting the observed displacement curves. This was plotted (see \cite{git:mey}) against the corresponding force components, $F_x$ and $F_z$, derived from the shapes. Thereby, {\bf GMBH} generated displays somewhat resembling those shown in Fig.\ \ref{fig6}, 
although with significant fluctuations (and digitizing noise) in all three variables $V$, $F_x$ and $F_z$. In the cases reported, the parallel components, $F_\parallel = -F_x$, started close to the stall force $F_{\mbox{\scriptsize S}} \simeq 6.5$ pN and decreased more or less steadily with time after buckling, while the perpendicular components, $F_\perp = F_z$, increased from close to zero. In parallel, the velocities started from low values, close to stall, but increased, sometimes very steeply with changing force, and eventually, for $F_\perp$ and $F_\parallel$ in the range 2$\,$$-5$ pN, passed through {\em maxima} as high as 850 to 1100 nm/s. Since these velocities correspond or exceed the highest normally seen in standard single-bead assays under low loads $|F_x|\lesssim 1$ pN, {\bf GMBH} concluded that the primary action of a perpendicular load was to increase $V$.

The $(V,F_{x,z})$ plots in Fig.\ \ref{fig6}, which roughly mimic the {\bf GMBH} observations, were generated from the velocity contours shown in Fig.\ \ref{fig5} by {\em postulating} the dashed and dot-dashed force trajectories labelled A and B, respectively. While our $(V,F_{x,z})$ plots resemble those of {\bf GMBH}, they attain much lower maximal speeds. This is clearly a consequence of the facts (a) that, as normal, $V(F_x=F_z=0) \simeq 800$ nm/s and (b) that for $V\gtrsim 200$ nm/s the contours of $V$ in Fig.\ \ref{fig5} slope upwards to the right so that, contrary to the conclusion of {\bf GMBH}, the predominant effect of increasing $F_z$ is to {\em reduce} the velocity. It should also be remarked that the close to $45^\circ$ slope of the velocity contours for $F_x\gtrsim 0$ directly reflects the observations \cite{blo:asb,car:cro} that even large assisting loads do {\em not} result in any significant increase in $V$.

Before discussing possible reasons for the disagreement with the buckling experiments, we ask what form the corresponding trajectories {\em should} take in the $(F_x,F_z)$ plane. We will learn that the putative trajectories A and B in Fig.\ \ref{fig5} are {\em not} very plausible.

Now, the persistence length of a microtubule is about 6 mm \cite{how}. Thus in analyzing their buckling data \cite{git:mey}, {\bf GMBH} considered an MT as a stiff, uniform rod with a measured flexual rigidity $EI$ (where $E$ is the Young's modulus and $I$ is the moment of inertia of the cross-section). The theory of bending a uniform rod is well established \cite{lan:lif}; but we will sketch it briefly in order to understand how the force trajectory in the buckling experiment may be derived. This then enables one to calculate the velocity of the kinesin molecule as it moves along the MT under the resulting induced load. 

We will now assume that the MT is clamped at the origin of the $(x,z)$ plane, while the motor is located close to the $x$-axis, say at $(L_0,0)$, where $L_0$ is the initial distance along the MT to the binding site in the clamped position: see Fig.\ \ref{fig3}. Note that the coordinates $x$ and $z$ here differ from those introduced in Fig.\ \ref{fig1} where the $x$-axis was taken parallel to the MT axis. Similarly, we will always denote the parallel and perpendicular components of the force relative to the MT as $F_\parallel$ and $F_\perp$, respectively. 

When a two-component force, ${\bm F}=(F_\parallel,F_\perp)$, is exerted on the MT by the tethered kinesin at $(L_0,0)$, the shape of the buckled MT (considered as a uniform rod) satisfies the beam equation which can be written \cite{how}
 \begin{equation}
  \frac{d^2\theta}{ds^2} = -\beta^2 \sin [\theta(s) - \varphi_F] \hspace{0.2in} \mbox{with} \hspace{0.2in} \beta^2 = F/EI,  \label{eq.27}
 \end{equation}
where $s$ is the arc length measured along the MT from the clamped origin $(0,0)$ while $\theta(s)$ is the tangential angle of the rod at the point $[x(s),z(s)]$ with respect to the $x$ axis; in addition, $\varphi_F$ denotes the angle of the applied force ${\bm F}$ (with respect to the $x$ axis) at the location of the kinesin which defines the {\em pivot point}, while $F = \sqrt{F_\parallel^2 + F_\perp^2}$ is the magnitude of the buckling force.

Assuming that the clamping of the MT is tight and that no torque is exerted at the pivot point, the solution of this equation must satisfy the boundary conditions
 \begin{equation}
  \theta(s{\,=\,}0) = 0 \hspace{0.3in} (d\theta/ds)_{s=L} = 0,  \label{eq.28} 
 \end{equation}
where $L$ is the total (time varying) arc length to the pivot point from the origin at the clamped position. Furthermore, the pivot point is fixed at the initial location of kinesin, which itself does not move spatially, so that one must have
 \begin{eqnarray}
  x(s{\,=\,}L) & = & \int_0^L \cos\theta(s)ds = L_0, \nonumber \\
  z(s{\,=}L) & = & \int_0^L \sin\theta(s) = 0.  \label{eq.29}
 \end{eqnarray}

Now, if the angle $\theta(s)$ remains small --- as it will when the MT starts to buckle --- one may accept the approximations $(dx/ds)=\cos\theta\approx 1$ and $(dz/ds)=\sin\theta\approx\theta$ which lead to
 \begin{equation}
  \theta \approx (dz/dx), \hspace{0.1in} (d^2\theta/ds^2) \approx (d^3z/ds^3) \approx (d^3z/dx^3).  \label{eq.30}
 \end{equation}
Expanding (\ref{eq.27}) up to first order then yields the linear equation
 \begin{equation}
  \frac{d^3 z}{dx^3} + \beta^2 \frac{dz}{dx} = \beta^2\varphi_F,  \label{eq.31}
 \end{equation}
for the displacement $z(x)$ of the MT from the $x$ axis. This is subject to the boundary conditions $z(0)=z^\prime (0) = z^{\prime\prime}(0) = z(L_0) = 0$ which then give
 \begin{equation}
  z(x) = \varphi_F {\bm (}x-L_0 + L_0\cos\beta x - \beta^{-1}\sin\beta x{\bm )}.  \label{eq.32}
 \end{equation}

But by (\ref{eq.29}) this form implies that the buckling force, given by $F=\beta^2 EI$, must satisfy the equation
 \begin{equation}
  \tan\beta L_0 = \beta L_0 \simeq 4.493.  \label{eq.33}
 \end{equation}
Note that in this small-$\theta$ approximation the magnitude of the force remains constant at $F^\ast \simeq EI(4.493/L_0)^2$ as the rod initially buckles. This in turn leads to a force trajectory in the $(F_x,F_z)$ plane that is just a semicircle of radius $F^\ast$. Of course, the small-angle approximation fails when $F_\perp$ increases.

Exact solutions to the nonlinear equation (\ref{eq.27}) can be expressed via an elliptic integral \cite{how,git:mey} as
 \begin{equation}
  \beta s = \int_{\phi_0}^{\phi(s)} \frac{d\phi}{\sqrt{1-k^2\sin^2\phi}},  \label{eq.34}
 \end{equation}
in which the modulus $k$ and elliptic angle $\phi(s)$ are related by
 \begin{equation}
  k\sin\phi(s) = \sin\mbox{$\frac{1}{2}$}[\theta(s)-\varphi_F],  \hspace{0.1in} k^2 = \sin^2\mbox{$\frac{1}{2}$}(\theta_L - \varphi_F),  \label{eq.35}
 \end{equation}
while $\theta_L \equiv \theta(s{\,=\,}L)$. Note that the solution represented by (\ref{eq.34}) already satisfies the boundary conditions (\ref{eq.28}). However, the initial elliptic angle $\phi_0$ and the modulus $k$ must be found so as to reproduce the correct pivot relations (\ref{eq.29}) while $\beta\propto\sqrt{F}$ determines the scale of the arc length. The parallel and perpendicular force components for an MT buckled to a contour length $L> L_0$ are then given by
 \begin{eqnarray}
  F_\parallel & = & F\cos(\theta_L - \varphi_F) = -F(1-2k^2),  \label{eq.36} \\
  F_\perp & = & F\sin(\theta_L - \varphi_F) = 2Fk\sqrt{1-k^2}.  \label{eq.37}
 \end{eqnarray}

Finally, for a given $L > L_0$, one may integrate (\ref{eq.29}) numerically using (\ref{eq.34}) and (\ref{eq.35}) to obtain the actual trajectory of MT buckling in the $(F_x,F_z)$ plane. The dotted curve in Fig.\ \ref{fig5} represents such a trajectory when the initial parallel force is set equal to the stall force $F_{\mbox{\scriptsize S}} \simeq 5.9$ pN. For small $F_\perp = F_z$, the trajectory approaches the circle described by the small-angle approximation.

\begin{figure}[ht]
\vspace{-0.8in}
\centerline{\psfig{figure=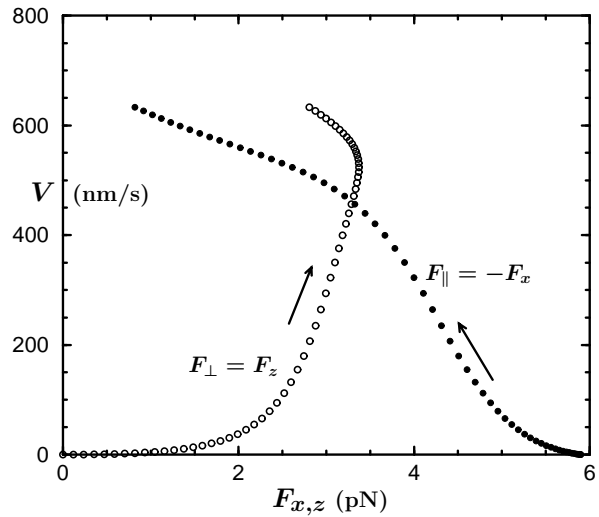,width=3.5in,angle=0}}
\vspace{-1.0in}
\caption{A velocity-force plot for a microtubule buckling experiment, as in Fig.\ \ref{fig6}, but based on a force trajectory calculated using theory for the bending of a uniform slender rod and shown as a dotted curve in Fig.\ \ref{fig5}.  \label{fig7}}
\end{figure}

The $(V,F_{x,z})$ buckling plot following from the calculated force trajectory (dotted curve in Fig.\ \ref{fig5}) is presented in Fig.\ \ref{fig7}. 
Evidently, this differs significantly from the results of {\bf GMBH} and from the postulated forms shown in Fig.\ \ref{fig6}. Indeed, the key experimental finding, namely, that the velocity, although noisy, exhibits a maximum above 850 nm/s at intermediate force levels is not reproduced: on the contrary one finds $V\lesssim 650$ nm/s in this region. On the other hand, the theory presented by {\bf GMBH}, which simplifies the functional form of the velocity to a linear expression in the force (see \cite{git:mey} Fig.\ 9c) generates qualitatively similar behavior to Fig.\ \ref{fig7}. Thus $F_\parallel$ falls steadily while $V$ always increases and $F_\perp$ passes through a maximum: see \cite{git:mey} Fig.\ 10.

In calculating the force trajectory for buckling an MT, we assumed that the clamping is tight and that no torque is applied by the motor protein at the pivot point. These assumptions seem reasonable based on the fits to the experimental data presented in \cite{git:mey} Fig.\ 5. Nevertheless, as one can detect in this figure, the experimental data, especially at the beginning of the buckling event, show small deviations from the fitted curve near the pivot point; indeed, the microtubule appears to be slightly bent at the pivot point. This suggests that the motor protein may actually exert a torque on the MT which might lead to a rather different force trajectory.

Other possible sources of uncertainty in the experiments and their interpretation were discussed critically by {\bf GMBH}, including clamp looseness, misalignment of the motor and the axis of clamping ($z=0$ in Fig.\ \ref{fig3}), protofilament number variations and slight, preformed bends in the microtubules, etc. Nevertheless, it is difficult to understand how the large maximal velocities observed, exceeding 900 nm/s, could be significantly in error. If this conclusion stands --- and, clearly, experiments in which $F_z$ and $F_x$ can be controlled more directly and with fewer uncertainties are to be desired --- a satisfactory explanation remains to be found. It is certain that the interaction of the heads, i.e., the motor domains with the microtubule plays a crucial role in kinesin motility: indeed, a recent normal mode analysis of simple protein network models \cite{zhe:don} highlights this feature as a distinction separating kinesin from standard myosin and the F1 ATPase motor. It seems possible, therefore, that the stressed state of the microtubule in the buckling experiments could be a prime cause of velocity enhancement.
\newpage
\section{Tether Linkage under Load}
\label{sec6}

Underlying our analysis \cite{fis:kim,kim:fis,fis:kim2} of the Block {\em et al.} experiments imposing reverse and transverse loads on kinesin \cite{lan:asb,blo:asb} are the expressions embodied in Models I and II for the perpendicular force, $F_z = {\mathcal F}_z (F_x)$, induced via the tether and bead by the longitudinal (or parallel) load: see (\ref{eq.24}) and (\ref{eq.25}) in Sec.\ \ref{sec4}. It is clearly of interest to attempt to test these forms against experiments that examine the transmission of force in the bead-tether-motor-track assembly. To that end we may utilize data obtained by Svoboda and Block \cite{svo:blo} who studied the elasticity of the bead-kinesin-microtubule linkage.

\begin{figure}[ht]
\vspace{-0.8in}
\centerline{\psfig{figure=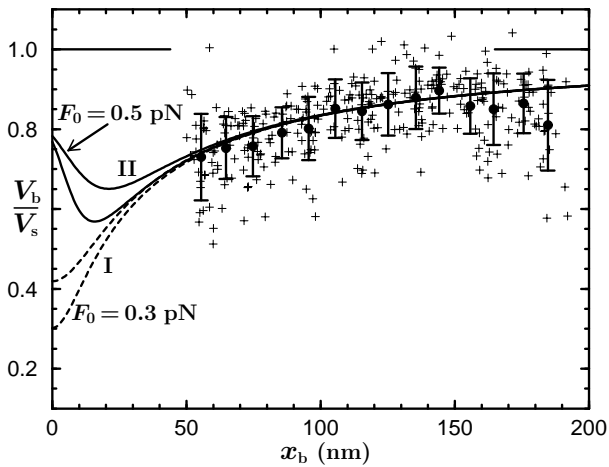,width=3.5in,angle=0}}
\vspace{-1.2in}
\caption{The ratio of bead velocity, $V_{\mbox{\scriptsize b}}$, to stage velocity, $V_{\mbox{\scriptsize s}}$, as a function of bead displacement, $x_{\mbox{\scriptsize b}}$, from the trap center, as observed obtained by Svoboda and Block \cite{svo:blo}. The solid circles are averages of the raw data (pluses) using a 10 nm bin size. The dotted lines are fits of Model I for the relation $F_z = {\mathcal F}_z(F_x)$, to the solid circles with noise levels $F_{0}=0.3$ pN and 0.5 pN, while the solid lines are corresponding fits for Model II.  \label{fig8}}
\end{figure}

These authors measured the ratio of the velocity $V_{\mbox{\scriptsize b}}$ of a bead trapped by optical tweezers to the velocity $V_{\mbox{\scriptsize s}}$ of the stage to which the MT was attached, in the presence of the nonhydrolyzable ATP analog AMP-PNP. This produces a rigorlike association between the kinesin motor and the MT. The ratio may then be expressed in terms of elasticities as
 \begin{equation}
  V_{\mbox{\scriptsize b}}/V_{\mbox{\scriptsize s}} = K_{\mbox{\scriptsize mot}}/(K_{\mbox{\scriptsize mot}}+K_{\mbox{\scriptsize b}}),  \label{eq.38}
 \end{equation}
where $K_{\mbox{\scriptsize mot}}$ and $K_{\mbox{\scriptsize b}}$ are the stiffness of the bead-motor-track linkage and of the bead in the optical trap, respectively. The ratio was measured as a function of the displacement $x_{\mbox{\scriptsize b}}$ of the bead from the center of the trap. The individual observations at a laser power of 62.5 mW are presented in Fig.\ \ref{fig8}. 
Although the data are noisy owing to the Brownian motion of the bead and to linkage heterogeneity \cite{svo:blo}, the mean ratio rises slowly as the displacement increases from the smallest observed value at $x_{\mbox{\scriptsize b}} = 50$ nm: see the solid circles which have been calculated by averaging the data binned in intervals of size 10 nm. We may then ask how well our models fit these data.

To proceed, note that the longitudinal force, $F_x$, applied by the optical trap for a bead displacement $x_{\mbox{\scriptsize b}}$ satisfies
 \begin{equation}
  - F_x = K_{\mbox{\scriptsize b}}x_{\mbox{\scriptsize b}} = K_{\mbox{\scriptsize mot}}\Delta l_x, \label{eq.39}
 \end{equation}
where $\Delta l_x$ is the change under tension of the projection of the kinesin tether on to the MT or $x$ axis. By assuming, as is reasonable \cite{how}, that the kinesin tether is rigid, we have $\Delta l_x = l\sin|\Theta|$ where $l$ is the length of the tether while $\Theta$, here negative, is the angle of the tether from the vertical axis: see Fig.\ \ref{fig2}. After some algebra, one finds
 \begin{equation}
  \frac{V_{\mbox{\scriptsize b}}}{V_{\mbox{\scriptsize s}}} = \frac{x_{\mbox{\scriptsize b}}}{x_{\mbox{\scriptsize b}} + l\sin|\Theta|},  \hspace{0.4in} \sin\Theta = \frac{F_x}{\sqrt{F_x^2 + F_z^2}}.  \label{eq.40}
 \end{equation}

Finally, we may use (\ref{eq.39}) in combination with the models (\ref{eq.24}) and (\ref{eq.25}) for ${\mathcal F}_z(F_x)$, to predict the ratio as a function of $x_{\mbox{\scriptsize b}}$. For the data in Fig.\ \ref{fig8} one has $K_{\mbox{\scriptsize b}} = 0.03$ pN/nm \cite{svo:blo}. If we take $l = 40$ nm, the models I and II yield perfectly satisfactory (if not very informative) fits to the data as evident from the solid and dashed lines in the figure. Note the two values assumed for the force fluctuation $F_0$ in (\ref{eq.24}) and (\ref{eq.25}). The tether length fitted here is shorter than what would be expected for a free kinesin molecule on the basis of photomicrographs \cite{how}, say, $l\lesssim 60$ nm. However, when a bead is chemically bound to a kinesin molecule it seems likely that some length of the tether may also be attached to the surface of the bead. Thus there are no grounds here for questioning the adequacy of the tether-based models for ${\mathcal F}_z (F_x)$.

\section{Summary}

We have extended the previous simple mechanochemical kinetic models for motor protein motion \cite{fis:kol,fis:kol2,kol:fis} to accommodate a three-dimensional free energy landscape for the point of attachment of the tether to the body of a motor that moves under a vector load ${\bm F}$. The load-dependence of the various forward and reverse rates describing the mechanochemical progress of the motor as it takes a single overall forward step of size $d$, are embodied, to first order in ${\bm F}$, in a set of load distribution vectors ${\bm \theta}_j^+$ and ${\bm \theta}_j^-$. These, in turn, relate directly to the forward {\em substeps}, $d_j (\leq d)$, taken by the motor along its track and also serve to localize the corresponding intermediate transition states. In quadratic order in the components $(F_x,F_y,F_z)$ of ${\bm F}$, a set of compliance matrices, ${\bm \eta}_j^+$ and ${\bm \eta}_j^-$, comes into play.

Even when perpendicular (or vertical) force components, $F_z$, are not purposefully imposed on a motor, as in standard single-bead optical-trap assays, consideration of the geometry of the bead-tether-motor-track configuration demonstrates that loads with $F_z > 0$ are {\em induced}. Furthermore, in switching between resistive and assisting loads (parallel to the track) accounting for these vertical components proves imperative.

The general analysis has been used to reconsider the microtubule buckling experiments of Howard and coworkers \cite{git:mey} in which the values of $F_z$ were measured. To that end the approach was first applied to the recent experiments of Block {\em et al.} \cite{blo:asb} in which, in particular, assisting loads imposed on kinesin were found to have little if any accelerating effect. The resulting fits provide velocity contours in the $(F_x,F_z)$ plane for kinesin under specified [ATP]: these are needed to address the buckling data of Howard and collaborators. Incidentally, however, the analysis also indicates that, on initially binding ATP, a kinesin motor ``crouches,'' i.e. moves downwards closer to the microtubule, prior to hydrolyzing ATP and stepping directly forward by $\sim\,$$8$ nm. Further and more detailed discussion of the Block {\em et al.} experiments on kinesin \cite{blo:asb} is provided elsewhere \cite{kim:fis,fis:kim2}; but, as shown here, the associated modelling of the transmission of tension via the bead-tether-kinesin-microtubule linkage is consistent with previous measurement by Svoboda and Block \cite{svo:blo}.

On the basis of the buckling experiments \cite{git:mey} Howard and coworkers concluded that a perpendicular force $F_z = 3\,$-5 pN, together with a longitudinal resisting force, $|F_x|$, of similar magnitude, results in kinesin velocities in excess of 900 nm/s. Our analysis does not substantiate this inference suggesting instead that velocities no larger than say 600 nm/s or so, should have been observed. A resolution of this challenging discrepancy might reside in the influence of microtubule stress or curvature on kinesin motility.

\acknowledgements
We are grateful to Steven Block for providing us with his laboratory's experimental data for kinesin. Interactions with him, Jonathon Howard, Matthew Lang and Anatoly Kolomeisky have been appreciated. The support of the National Science Foundation (through Grant No.\ CHE 03-01101) is gratefully acknowledged.



\begin{thebibliography}{00}
  \bibitem{how} Howard J 2001 {\em Mechanics of Motor Proteins and the Cytoskeleton} (Sunderland, Mass: Sinauer Assoc)
  \bibitem{koj:mut} Kojima H, Muto E, Higuchi H and Yanagida T 1997 {\em Biophys.\ J.} {\bf 73} 2012
  \bibitem{vis:sch} Visscher K, Schnitzer M J and Block S M 1999 {\em Nature} {\bf 400} 184
  \bibitem{sch:vis} Schnitzer M J, Visscher K and Block S M 2000 {\em Nature Cell Biol.} {\bf 2} 718
  \bibitem{fis:kol} Fisher M E and Kolomeisky A B 1999 {\em Proc.\ Natl.\ Acad.\ Sci.\ USA} {\bf 96} 6597
  \bibitem{fis:kol2} Fisher M E and Kolomeisky A B 2001 {\em Proc.\ Natl.\ Acad.\ Sci.\ USA} {\bf 98} 7748
  \bibitem{kol:fis} Kolomeisky A B and Fisher M E 2003 {\em Biophys.\ J.} {\bf 84} 1642
  \bibitem{cop:pie} Coppin C M, Pierce D W, Hsu L and Vale R D 1997 {\em Proc.\ Natl.\ Acad.\ Sci.\ USA} {\bf 94} 8539
  \bibitem{lan:asb} Lang M J, Asbury C L, Shaevitz J W and Block S M 2002 {\em Biophys.\ J.} {\bf 83} 491
  \bibitem{blo:asb} Block S M, Asbury C L, Shaevitz J W and Lang M J 2003 {\em Proc.\ Natl.\ Acad.\ Sci.\ USA} {\bf 100} 2351
  \bibitem{car:cro} Carter N J and Cross R A 2005 {\em Nature} {\bf 435} 308
  \bibitem{svo:blo} Svoboda K and Block S M 1994 {\em Cell} {\bf 77} 773
  \bibitem{git:mey} Gittes F, Meyh\"{o}fer E, Baek S and Howard J 1996 {\em Biophys.\ J.} {\bf 70} 418
  \bibitem{fis:kim} Fisher M E and Kim Y C 2004 {\em Biophys.\ J.} {\bf 86} 527a
  \bibitem{kim:fis} Kim Y C and Fisher M E 2005 {\em Biophys.\ J.} {\bf 88} 649a
  \bibitem{nis:mut} Nishiyama M, Muto E, Inoue Y, Yanagida T and Higuchi H 2001 {\em Nature Cell Biol.} {\bf 3} 425
  \bibitem{lei:hus} Leibler S and Huse D A 1993 {\em J.\ Cell Biol.} {\bf 121} 1357
  \bibitem{jul:adj} J\"{u}licher F, Adjari A and Prost J 1997 {\em Rev.\ Mod.\ Phys.} {\bf 69} 1269
  \bibitem{kol:wid} Kolomeisky A B and Widom B 1998 {\em J.\ Stat.\ Phys.} {\bf 93} 633
  \bibitem{bus:kel} Bustamante C, Keller D and Oster G 2001 {\em Acc.\ Chem.\ Res.} {\bf 34} 412
  \bibitem{sch:cla} Schief W R, Clark R H, Crevenna A H and Howard J 2004 {\em Proc.\ Natl.\ Acad.\ Sci.\ USA} {\bf 101} 1183
  \bibitem{kol:fis2} Kolomeisky A B and Fisher M E 2000 {\em Physica A} {\bf 279} 1; {\em J.\ Chem.\ Phys.} {\bf 113} 10867
  \bibitem{yil:tom} Yildez A, Tomishige M, Vale R D and Selvin P R 2004 {\em Science} {\bf 303} 676
  \bibitem{klu:hoe:gil} Klumpp L M, Hoenger A and Gilbert S P 2004 {\em Proc.\ Natl.\ Acad.\ Sci.\ USA} {\bf 101} 3444
  \bibitem{kol:phi} Kolomeisky A B and Philips H III 2005 arXiv:cond-mat/0503169
  \bibitem{fis} Fisher D S 1997 {\em Physica D} {\bf 107} 204
  \bibitem{sac:lec} Sachs F and Lecar H 1991 {\em Biophys.\ J.} {\bf 59} 1143
  \bibitem{fis:kim2} Fisher M E and Kim Y C [to be published]
  \bibitem{lan:lif} Landau L D and Lifshitz E M 1986 {\em Theory of Elasticity} 3rd Ed (Oxford: Pergamon Press)
  \bibitem{zhe:don} Zheng W and Doniach S 2003 {\em Proc.\ Natl.\ Acad.\ Sci.\ USA} {\bf 100} 13253
\end{thebibliography}
\end{document}